# Large-Scale Fabrication of RF MOSFETs on Liquid-Exfoliated MoS$_2$

Kuanchen Xiong[#1], Lei Li[#], Asher Madjar[#], James C. M. Hwang[#], Zhaoyang Lin[*], Yu Huang[*], Xiangfeng Duan[*],
Alexander Göritz[+], Matthias Wietstruck[+], and Mehmet Kaynak[+]

[#]Lehigh University, Bethlehem, Pennsylvania 18015 USA
[*]University of California, Los Angeles, California 90095 USA
[+]IHP, Im Technologiepark 25, 15236 Frankfurt (Oder), Germany
[1]kux214@lehigh.edu

*Abstract* — For the first time, thousands of RF MOSFETs were batch-fabricated on liquid-exfoliated MoS$_2$ below 300 °C with nearly 100% yield. The large-scale fabrication with high yield allowed the *average* performance instead of the *best* performance to be reported. The DC performance of these devices were typical, but the RF performance, enabled by buried gates and on the order of 100 MHz, was reported for the first time for liquid-exfoliated MoS$_2$. To resolve the dilemma of thin vs. thick films, gate recess was used on 20-nm thick films to improve the gate control while keeping the contact resistance lower than that on 10-nm films. These innovations may enable thin-film transistors to operate in the microwave range.

*Index Terms* — Electrochemical process, CMOS process, MOSFET, semiconductor device manufacture, semiconductor nanostructures, thin film transistors, wafer scale integration.

## I. INTRODUCTION

Two-dimensional atomic-layered materials such as 2H-MoS$_2$ have high charge carrier mobility without epitaxy requirement. They are therefore promising materials for thin-film transistors to operate in the microwave range, well beyond that of current-generation InGaZnO devices. To date, most MoS$_2$ transistors have been individually crafted by direct-write electron-beam lithography on micrometer-size flakes mechanically exfoliated from bulk crystals. Recently, scores of MOSFETs and small-scale integrated circuits were fabricated on large-area MoS$_2$ synthesized by chemical vapor deposition [2]. However, the process required a thermal budget on the order of 1000 °C, which is incompatible with typical thin-film processes.

Table 1. MOSFETs Fabricated on Liquid-Exfoliated MoS$_2$

| Year | MoS$_2$ Flake Thick. nm | MoS$_2$ Flake Length μm | FET Channel Thick. nm | FET Channel Length μm | Coverage | Gate | Mobility cm$^2$/Vs | On/Off Ratio | Ref. |
|---|---|---|---|---|---|---|---|---|---|
| 2011 | 10 | 1 | 100 | 1 | Low | Subs. | 0.1 | 10 | [3] |
| 2011 | 1 | 1 | 1 | 1 | Low | Subs. | 0.001 | --- | [4] |
| 2013 | 10 | 10 | 10 | 1 | Low | Ion | 1 | 100 | [5] |
| 2014 | 1 | 10 | 1 | 10 | Low | Subs. | 1 | 10$^6$ | [6] |
| 2014 | 10 | 0.1 | 100 | 10 | High | Subs. | 10$^{-5}$ | 10 | [7] |
| 2017 | 10 | 1 | 1000 | 100 | High | Ion | 0.1 | 100 | [8] |
| Present | 10 | 1 | 10 | 1 | High | Buried | 0.01 | 10 | This Work |

By contrast, liquid-exfoliated and dispensed MoS$_2$ can be applied near room temperature, in large area, and at low cost, making it attractive for thin-film processes. Over the past decade, there have been many attempts to demonstrate MOSFETs on liquid-exfoliated MoS$_2$ as listed in Table 1. However, they typically exhibited either high performance on thin film with low coverage, or low performance on thick film with high coverage. Additionally, they used either the entire substrate or an ion liquid as the gate electrode, rendering them inoperable at radio frequencies. To resolve the dilemma of thin vs. thick films, we recessed the gate following the established practice on compound semiconductors [9] and recent attempts on black phosphorous [10], [11]. To enable RF operation, we fabricated submicron buried gates with low parasitic inductance and capacitance. These innovations are described in the following.

## II. DEVICE FABRICATION

Thousands of MoS$_2$ MOSFETs were batch-processed through three major steps: (A) formation of buried gates, (B) spin-coating of MoS$_2$, and (C) definition of active and contact regions. To facilitate the gate-recess experiment, although step A was carried out on a 200-mm Si wafer, the wafer was subsequently diced into 25 mm × 15 mm chips before steps B and C.

A one-mask photolithography process was used to form Al gates buried in SiO$_2$, before deposition of gate oxide and MoS$_2$ (Fig. 1). This way, gate oxide could be deposited at relatively high temperature without damaging MoS$_2$. To this end, state-of-the-art CMOS processes are capable of not only submicron gate and high-quality gate oxide, but also flat surface through chemo-mechanical polishing to ensure smooth dispense of MoS$_2$. Specifically, the back-end-of-line process of the SG13S foundry process by IHP Microelectronics with a thermal budget of 450 °C was chosen. Using a 200-mm high-resistivity (10 kΩ·cm) Si wafer, approximately sixty 25 mm × 15 mm chips were fabricated, with each chip containing approximately 1500 RF-probable MOSFETs. Each MOSFET has two buried gates with a total gate width of approximately 10 μm. The source-gate spacing, gate length, and gate-drain spacing were approximately 0.4 μm, 0.6 μm, and 0.4 μm, respectively. The gate thickness was approximately 0.5 μm. Following chemo-mechanical polishing, Al$_2$O$_3$ gate oxide approximately 30-nm thick was deposited by atomic layer deposition (ALD) at 250 °C. Fig. 2(a) shows a MOSFET at this process step.

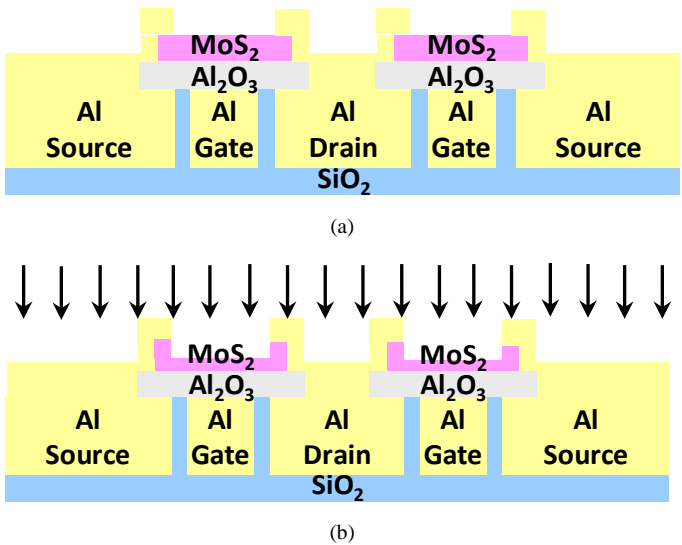

Fig. 1. Cross-section schematic (not to scale) of a MoS$_2$ MOSFET (a) before and (b) after gate recess.

The isopropanol suspension of MoS$_2$ contained flakes with a thickness of 3–6 nm and a lateral size of 1−2 μm. The suspension was spun onto the chips of step A at 2000 rpm for 20 s, followed by annealing at 300 °C for 1 h in argon. The spin-coating process was repeated two and four times to obtain MoS$_2$ films with approximately 10- and 20-nm thicknesses, respectively, as confirmed by atomic force microscopy. The atomic force microscopy confirmed also that the films were continuous despite a roughness of approximately 5 nm, which resulted in grainy contrast under the optical microscope as shown in Fig. 2(b). The chip coated with 10-nm MoS$_2$ served as a control; the chip coated with 20-nm MoS$_2$ would undergo gate recessing.

A two-mask photolithography process was used to define the active and contact regions of all 1500 MOSFETs on each chip. To define the active region, MoS$_2$ was dry-etched by CHF$_3$/O$_2$ whereas Al$_2$O$_3$ was wet-etched by buffered HF. Source and drain contacts were formed by electron-gun evaporated Ti and Al with thicknesses of 10 nm and 490 nm, respectively. Gate recessing was performed by dry etching with CHF$_3$/O$_2$ for 50 s using the source and drain contacts as self-aligned masks.

## III. Result and Discussion

Random sampling confirmed that, before gate recessing, nearly all 1500 MOSFETs fabricated on each chip had source-drain conduction, whether the chip was coated with 10-nm or 20-nm MoS$_2$. However, whereas devices on 10-nm MoS$_2$ exhibited effective gate control, devices on 20-nm MoS$_2$ exhibited no gate modulation up to a gate-source voltage of 10 V, presumably because the channel was too thick. Correspondingly, the total source-drain resistance was reduced from 28 ± 8

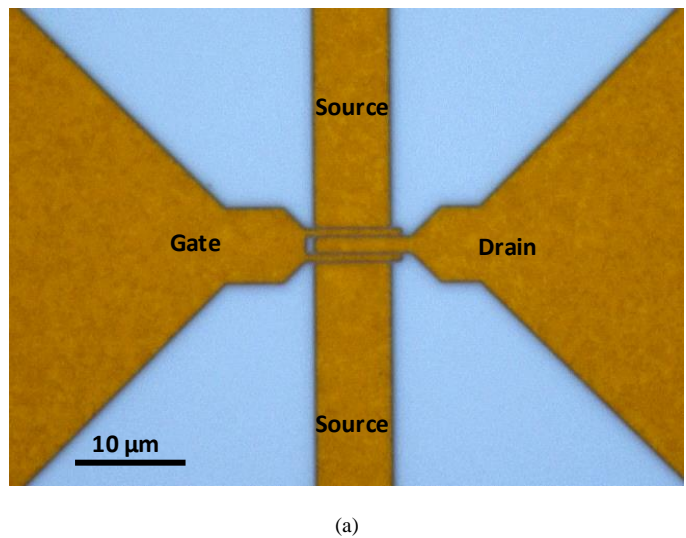

(a)

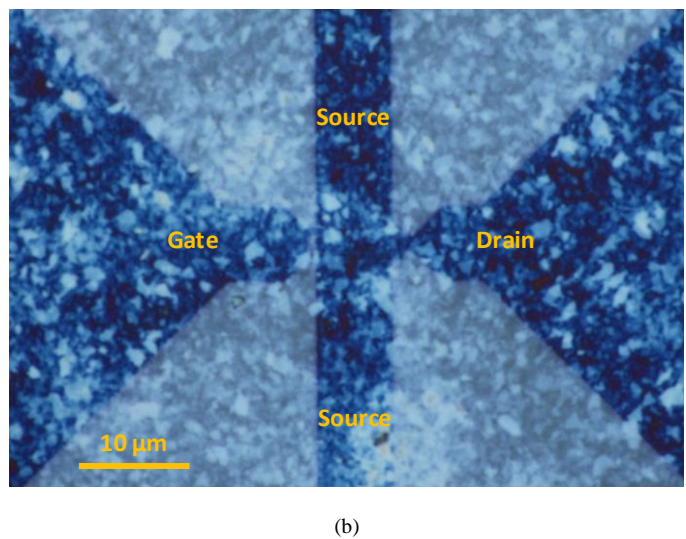

(b)

Fig. 2. Micrographs of a MOSFET (a) before and (a) after spin-coating of 10-nm-thick MoS$_2$.

MΩ·μm for devices on 10-nm MoS$_2$ to 20 ± 10 MΩ·μm for devices on 20-nm MoS$_2$. (The reduction is less than 50% because the source-drain resistance is presently dominated by the contact resistance.) Devices on 20-nm MoS$_2$ exhibited effective gate control only with gate recess. However, the gate was over-recessed and only 90 of the 1500 devices survived, with the yield dropping from 100% to approximately 6%. Visual inspection confirmed that only 167 devices had channel material left after gate recessing. Thus, although the present experiment demonstrated the benefit of gate recessing in improving gate control and reducing contact resistance, much process development and control are needed. For example, gate recessing could be done with wet etching instead of dry etching to avoid damaging the channel material. Historically, gate recessing has been the most critical process step for compound semiconductor devices [9].

For the chip coated with 10-nm MoS$_2$, Fig. 3(a) shows the transfer characteristics of a typical MOSFET. The drain current of approximately 0.15 µA/µm, the field-effect mobility of approximately 0.03 cm$^2$/Vs, and the on/off current ratio of approximately 12 are typical of MOSFETs fabricated on liquid-exfoliated MoS$_2$ as listed in Table I. (The field-effect electron mobility $\mu_{FE}$ is calculated according to $\mu_{FE} = g_M L / C_{OX} V_{DS} W$, where $g_M$ is the peak transconductance, $L$ and $W$ are the channel length and width, respectively, $C_{OX}$ is the oxide capacitance, and $V_{DS}$ is the drain-source voltage.) However, the de-embedded (from large probe pad capacitances [12]) forward-current cutoff frequency $f_T$ of approximately 38 MHz and the maximum oscillation frequency of $f_{MAX}$ of approximately 49 MHz (Fig. 4) are the first time to be reported for any MOSFET fabricated on liquid-exfoliated MoS$_2$. Although the $f_T$ and $f_{MAX}$ are far from the microwave range, they are already comparable or better than that of current-generation InGaZnO devices. Detailed analysis showed that presently $f_T$ and $f_{MAX}$ are mainly limited by large parasitic resistances associated with both the source contacts and the source-gate access region. These parasitic resistances also prevented the drain current from saturating below 3 V. These parasitic resistances may be reduced by a shorter source-gate distance and a more optimum gate-recess profile. Nevertheless, Fig. 3(a) shows also that the gate leakage current is less than 1 nA/µm, which attests to the high-quality gate oxide afforded by the buried-gate configuration.

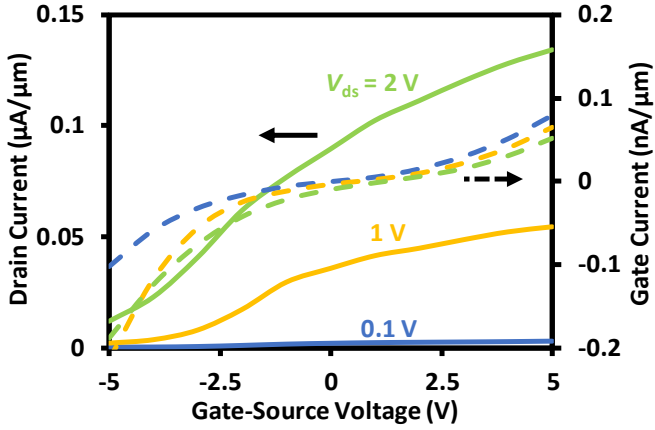

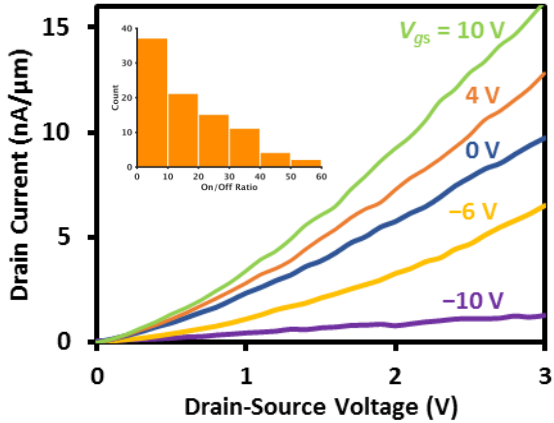

Fig. 3. (a) Transfer characteristics of a typical MOSFET fabricated on 10-nm MoS$_2$ *without* gate recess, and (b) drain characteristics of a typical MOSFET fabricated on 20-nm MoS$_2$ *with* gate recess. Inset in (b) shows histograhm of on/off current ratio.

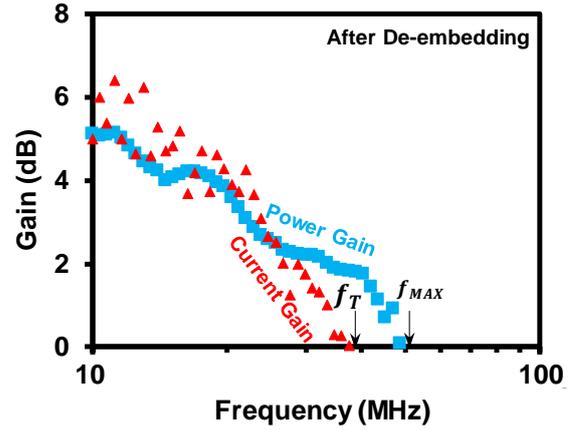

Fig. 4. De-embedded forward current gain cutoff frequency $f_T$ and maximum frequency of oscillation $f_{MAX}$ of a typical MOSFET fabricated on 10-nm liquid-exforliated MoS$_2$.

For the chip coated with 20-nm MoS$_2$, Fig. 3(b) shows that gate recess improves the on/off current ratio, but reduces the current capacity, presumably due to damage by dry etching. The inset shows that the on/off ratio of the 90 surviving devices is 18 ± 12 with the maximum being 60. Correspondingly, the threshold is −9 V ± 6 V. The present large-scale fabrication allows *average* characteristics to be reported, whereas others in Table I report *best* characteristics.

## IV. CONCLUSION

For the first time, thousands of RF MOSFETs were batch-fabricated on liquid-exfoliated MoS$_2$ with nearly 100% yield. The large-scale fabrication allowed the *average* performance instead of the *best* performance to be reported. The DC performance of these devices were typical of that of MOSFETs fabricated on liquid-exfoliated MoS$_2$, with drain current ~ 0.1 µA/µm, electron mobility ~ 0.01 cm$^2$/Vs, and on/off ratio ~ 10. However, the RF performance, on the order of 100 MHz, was reported for the first time. The RF performance was enabled by buried gates with low parasitic inductance and capacitance. Additionally, to resolve the dilemma of thin vs. thick films, gate recess was introduced to MoS$_2$ MOSFETs for the first time, which improved gate control and contact resistance. However, the present process is far from optimum for achieving the full potential of the material, which had been shown to result in orders-of-magnitude higher mobility and on/off ratio with a substrate gate. Therefore, with further process development and optimization, especially in reducing the parasitic resistances of the source contact and the source-gate access region, these innovations and results suggest that MoS$_2$ MOSFETs may enable thin-film transistors to operate in the microwave range.

ACKNOWLEDGMENT

This work was supported in part by the U.S. Office of Naval Research under Grant N00014-14-1-0653, the Air Force Office of Scientific Research and the National Science Foundation (NSF) under EFRI 2-DARE Grant 1433541 and Grant 1433459. This work was performed in part at the Cornell NanoScale Facility, a member of the National Nanotechnology Coordinated Infrastructure, which is supported by NSF under Grant ECCS-1542081.